\documentclass[12pt]{article}
\usepackage[dvips]{graphicx}
\usepackage{amssymb,amsmath,graphicx}

\def\lromn#1{\uppercase\expandafter{\romannumeral#1}}
\def\ds{\displaystyle}
\def\DM{\tilde{\chi}^0}
\def\CP{\tilde{\chi}^-}
\def\CPC{\tilde{\chi}^+}

\begin{document}
  
\begin{titlepage}

\begin{center}

\hfill KEK-TH-1111 \\
\hfill \today

\vspace{1cm}
{\large\bf Non-perturbative Effect\\ on Thermal Relic Abundance of Dark Matter}
\vspace{1.5cm}

{\bf Junji Hisano}$^{(a)}$,
{\bf Shigeki Matsumoto}$^{(b)}$,
{\bf Minoru Nagai}$^{(a)}$,
{\bf Osamu Saito}$^{(a,b)}$,
and
{\bf Masato Senami}$^{(a)}$

\vskip 0.15in

{\it
$^{(a)}${ICRR, University of Tokyo, Kashiwa 277-8582, Japan} \\
$^{(b)}${Theory Group, KEK, Oho 1-1 Tsukuba, Ibaraki 305-0801, Japan} \\
}

\vskip 1in

\abstract{ We point out that thermal relic abundance of the dark
  matter is strongly altered by a non-perturbative effect called the
  Sommerfeld enhancement, when constituent particles of the dark
  matter are non-singlet under the SU(2)$_L$ gauge interaction and
  much heavier than the  weak gauge bosons. 
  Typical candidates for such dark matter particles are the heavy
  wino- and higgsino-like neutralinos.  We investigate the
  non-perturbative effect on the relic abundance of dark matter for
  the wino-like neutralino as an example.  We show that its thermal
  abundance is reduced by 50\% compared to the perturbative result.
  The wino-like neutralino mass consistent with the observed dark
  matter abundance turns out to be 2.7~TeV $\lesssim m \lesssim$
  3.0~TeV.}

\end{center}
\end{titlepage}
\setcounter{footnote}{0}

\section{Introduction}

Precise measurements of cosmological parameters have achieved amazing
progress in recent years. In particular, the observation of cosmic
microwave background anisotropies by the Wilkinson Microwave
Anisotropy Probe (WMAP) \cite{WMAP} confirmed that non-baryonic dark
matter amounts to 20\% of the energy of our universe.  The existence
of dark matter forces us to consider physics beyond the standard model
(SM) for its constituent, because the SM has no candidate for the dark
matter.

Many models beyond the SM have been proposed for providing dark matter
candidates.  Models involving weakly interacting massive particles
(WIMPs), such as supersymmetric models and universal extra dimension
(UED) models \cite{UED,UED2}, have an advantage over other models,
since the WIMPs can explain the observed abundance naturally in the
thermal relic scenario \cite{UED2,susydm,UEDabundance}.  Comparison of
the predicted thermal abundance with observations is a powerful tool
to constrain those models. It is now important for discussion of new
physics signatures at collider experiments such as the LHC.

In the thermal relic scenario, the annihilation cross sections of dark
matter play a central role in evaluation of the abundance.
Perturbative calculation of the cross sections is sufficient for it in
usual cases.  However, a non-perturbative effect on the cross
sections has to be included, when dark matter particles are
non-singlet under the SU(2)$_L$ interaction and much heavier than the
weak gauge bosons. The weak interaction is not a short-distance force,
but rather a long-distance one for non-relativistic two-bodies states
of such heavy particles. The wave functions for the two-bodes states
are modified from the plane waves by the interaction, and the
annihilation cross sections are affected. The dark matter particles
are non-relativistic at the freeze-out temperature.  The attractive
channels due to the weak force enhance the annihilation cross
sections, compared with the perturbative ones. This effect is called
the Sommerfeld enhancement, which was found in inelastic reactions
between non-relativistic charged particles, historically.

In this letter, we point out that the non-perturbative effect strongly
alters the relic abundance of dark matter compared to that in
perturbative calculation. We investigate the non-perturbative effect
quantitatively in the wino-like neutralino dark matter as an example,
and show that the abundance is reduced by 50\% due to the effect. We also
discuss non-perturbative effects on other dark matter candidates at
the end of this letter.

\newpage

\section{Sommerfeld enhancement in wino-like neutralino dark matter annihilation}

Winos are the SU(2)$_L$ gauginos in the supersymmetric standard model,
and they are an SU(2)$_L$ triplet. They are mixed with higgsinos,
superpartners of the Higgs bosons, after the electroweak symmetry
breaking. The neutral components in the mass eigenstates are called
neutralino. The mixing with higgsinos is small for heavy winos,
since it is suppressed by the electroweak scale.
Thus, the wino-like neutralino ($\DM$) is highly degenerate
with its charged SU(2)$_L$ partner ($\CP$) in mass when their masses are
heavy enough. For the wino-like neutralino mass ($m$) of
the order of 1~TeV, the mass difference between the neutralino and
its SU(2)$_L$ partner is dominated by a radiative correction
\cite{Hisano:2004ds}.  It is calculated as $\delta m \simeq$ 0.17 GeV,
which is used for numerical calculations in this letter.

There are four processes related to the calculation of the wino-like
neutralino relic abundance, and those are $\DM\DM$, $\CP\CPC$,
$\DM\CP$, and $\CP\CP$ annihilation.  We assume the CP conservation,
and cross sections of $\DM\CPC$ and $\CPC\CPC$ annihilation are the
same as those of $\DM\CP$ and $\CP\CP$, respectively.  Furthermore,
each process is decomposed into two ones with $S = 0$ and $S = 1$,
where $S$ is the total spin of the two-bodies system at the initial
state.  In $\DM\DM$ and $\CP\CP$ annihilation, the $S = 1$ processes are
forbidden at the $s$-wave annihilation.  After all, processes we have
to consider are $\DM\DM$, $\CP\CPC$, $\DM\CP$, $\CP\CP$ annihilation
with $S = 0$, and those of $\CP\CPC$, $\DM\CP$ with $S = 1$.

While gamma rays from the wino-like neutralino annihilation in the
Galactic center \cite{Hisano:2004ds,Hisano:2003ec} and anti-particles
fluxes from that in Galactic Halo \cite{Hisano:2005ec} are evaluated including
non-perturbative effects, the relic abundance of the neutralino have
been calculated only within a perturbative method.  However, as we
will see later, the non-perturbative effect can significantly alter
cross sections relevant to the abundance.

Two-bodies effective Lagrangian, describing relative motion between
two particles in the two-bodies system, is useful to evaluate the
non-perturbative effect in the annihilation. The Lagrangian has a
following form,
\begin{eqnarray}
  {\cal L}_{\rm eff}
  =
  \int d^3 r~
  \sum_{S_Z}\Phi^\dagger_{S_Z}(x,\vec{r})
  \left[
    i\partial_{x^0} + \frac{\nabla_x^2}{4m} + \frac{\nabla_r^2}{m}
    -
    V(r)
    +
    2i \Gamma \delta(\vec{r})
  \right]
  \Phi_{S_Z}(x,\vec{r})~,
\label{twob}
\end{eqnarray}
where $x$ is the center of mass coordinate for the two-bodies system,
while the relative one is represented by $\vec{r}$.  The index $S_Z$
is for the $z$-th component of the total spin $S$.  The potential term
$V(r)$ describes forces acting between two particles in the system.
The absorptive part $\Gamma$ stands for the annihilation of the
two-bodies system. For  derivation of the effective Lagrangian and
following evaluation of the annihilation cross section using it,
see Ref.~\cite{Hisano:2004ds}.

In the $\CP\CP$ annihilation process with $S = 0$, the potential
$V(r)$ and absorptive part $\Gamma$ in Eq.~(\ref{twob}) turn out to be
\begin{eqnarray}
 V(r)
 =
 \frac{\alpha}{r} + \alpha_2 c_W^2 \frac{e^{-m_Z r}}{r}~,
 \quad
 \Gamma
 =
 \frac{\pi\alpha_2^2}{2m^2}~,
\end{eqnarray}
where $\alpha$ is the fine structure constant, $\alpha_2 = g_2^2/4\pi$
is for the SU(2) gauge coupling constant, $m_Z$ is the $Z$ boson mass,
and $c_W = \cos\theta_W$ is for the weak mixing angle $\theta_W$. The
first term in the potential comes from the exchange of photons, while
the second one is from the exchange of $Z$ bosons.  In the calculation
of $\Gamma$, we consider only the final states of SM particles, and
neglect their masses, since they are light enough compared to the
wino-like neutralino we are discussing.

In both cases of the $\DM\CP$ annihilation processes with $S = 0$ and
$1$, $V(r)$ is induced from the exchange of $W$ bosons, and both cases
have the same form.  On the other hand, the absorptive part $\Gamma$
is different each other.  These are given by
\begin{eqnarray}
 V(r)
 =
 - \alpha_2 \frac{e^{-m_W r}}{r}~,
 \quad
 \Gamma_{(S = 0)} 
 =
  \frac{1}{2} \frac{\pi\alpha_2^2}{m^2}~,
 \quad
 \Gamma_{(S = 1)} 
 =
 \frac{25}{24} \frac{\pi\alpha_2^2}{m^2}~,
\end{eqnarray}
where $m_W$ is the $W$ boson mass. The potential and absorptive terms
in the $\CP\CPC$ annihilation with $S = 1$ are
\begin{eqnarray}
 V(r)
 =
 - \frac{\alpha}{r} - \alpha_2 c_W^2 \frac{e^{-m_Z r}}{r}~,
 \qquad
 \Gamma
 =
 \frac{25}{24} \frac{\pi \alpha_2^2}{m^2}~.
\end{eqnarray}

The $\DM\DM$ two-bodies system is mixed with $\CP\CPC$ state with $S = 0$, in
which mixing occurs through a $W$ boson exchange. Thus, the potential and
absorptive terms are written by $2 \times 2$ matrices as
\begin{eqnarray}
 {\bf V}(r)
 =
 \left(
  \begin{array}{cc}
   2\delta m - \ds\frac{\alpha}{r} - \ds\alpha_2 c_W^2 \frac{e^{-m_Z r}}{r}
   &
   -\sqrt{2} \alpha_2 \ds\frac{e^{-m_W r}}{r}
   \\
   -\sqrt{2} \alpha_2 \ds\frac{e^{-m_W r}}{r} 
   &
   0
  \end{array}
 \right),
 ~
 {\bf \Gamma}
 =  
 \frac{\pi \alpha_2^2}{2 m^2}
 \left(
  \begin{array}{cc}
   3 & \sqrt{2}
   \\
   \sqrt{2} & 2
  \end{array}
 \right).
\end{eqnarray}
Off-diagonal elements describe the transition between $\CP\CPC$ and $\DM\DM$
systems.

As seen in these potentials, all processes have attractive channels
except for that of $\tilde \chi ^- \tilde \chi ^-$.  The overlap
between the wave functions of the incident particles are increased
compared to the case without including the potentials, and it leads to
enhancement of the annihilation cross sections.

Once the two-bodies effective Lagrangian is obtained, annihilation
cross sections including the non-perturbative effect can be calculated
through the formula,
\begin{eqnarray}
 \sigma v
 =
 c
 \Gamma
 |A|^2~,
 \qquad
 A
 \equiv
 \int d^3r~
 e^{-i\vec{k}\cdot\vec{r}}
 \left(
  \frac{mv^2}{4}
  +
  \frac{\nabla^2}{m}
 \right)
 G(\vec{r},\vec{0})~,
 \label{formula}
\end{eqnarray}
where $c = 2$ for an annihilation of identical particles, otherwise
$c=1$.  The relative velocity between incoming particles is denoted by
$v$.  The Green function $G$ satisfies the equation of motion of the
effective Lagrangian,
\begin{eqnarray}
 \left[
  \frac{mv^2}{4}
  +
  \frac{\nabla^2}{m}
  -
  V(r) 
  +
  2i\Gamma\delta^3(\vec{r})
 \right]G(\vec{r},\vec{r}')
 =
 \delta^3(\vec{r} - \vec{r}')~.
 \label{EOM}
\end{eqnarray}
The boundary condition for the Green function is determined from
following conditions. First, the Green function is analytic at any
$\vec{r}$ and $\vec{r}'$ except the point $\vec{r} =
\vec{r}'$. Second, only out-going waves survive at large $|\vec{r} -
\vec{r}'|$.

In $\DM\DM$ and $\CP\CPC$ annihilation processes with $S = 0$, the Green
function has a $2\times2$ matrix form.
For this case, it has been found in Ref.~\cite{Hisano:2004ds,Hisano:2003ec}
that above formula is simply extended as
\begin{eqnarray}
 \sigma_i v
 =
 c_i
 \sum_{j,j'}
 {\bf A}_{ij}
 {\bf \Gamma}_{jj'}
 {\bf A}^*_{ij'}~.
\end{eqnarray}

The factor $|A|^2$ in Eq.~(\ref{formula}) is called the Sommerfeld
enhancement factor, and it includes the non-perturbative effect due to
the long-distance forces in $V(r)$. Note that if we neglect the
non-perturbative effect, the enhancement factor becomes one and the
annihilation cross section in Eq.~(\ref{formula}) coincides with the
result in a usual perturbative method.


\section{Effect of Sommerfeld enhancement on dark matter abundance}
 
Now we evaluate the thermal relic abundance of the wino-like
neutralino dark matter, including the non-perturbative effect. In the
evaluation we have to include coannihilation processes in addition to
the wino-like neutralino pair annihilation.  We use the method
developed in Ref.~\cite{Griest:1990kh,Gondolo:1990dk} for the
calculation of the relic abundance including coannihilation
effects. Under reasonable assumptions, the relic density obeys the
following Boltzmann equation,
\begin{eqnarray}
 \frac{dY}{dx}
 =
 -\frac{\langle\sigma_{\rm eff} v\rangle}{Hx}
 \left(1 - \frac{x}{3g_{*s}}\frac{dg_{*s}}{dx}\right)
 s\left(Y^2 - Y_{\rm eq}^2\right)~.
 \label{eq:Boltzmann}
\end{eqnarray}
The yield of the dark matter, $Y$, is defined as $Y = n/s$, where $n$ is the sum
of the number densities of $\DM$, $\CP$, and $\CPC$.
The variable, $x = m/T$, is the scaled inverse temperature of the universe.
The equilibrium abundance, $Y_{\rm eq}$, is given by
\begin{eqnarray}
 Y_{\rm eq}
 =
 \frac{45}{2\pi^4} \left(\frac{\pi}{8}\right)^{1/2}
 \frac{g_{\rm eff}}{g_{*}} x^{3/2} e^{-x}~,
\end{eqnarray}
where $g_{\rm eff}$ is the number of the effective degrees of freedom 
defined as
\begin{eqnarray}
 g_{\rm eff}(x)
 = 
 2 + 4(1 + \delta m/m)^{3/2} e^{-x \delta m/m}~.
 \label{eq:geff}
\end{eqnarray}
The entropy density $s$ and the Hubble parameter $H$ are given by
\begin{eqnarray}
 s
 =
 \frac{2\pi^2}{45} g_{*s}\frac{m^3}{x^3}~,
 \qquad
 H =  \left( \frac{g_*}{10} \right)^{1/2} \frac{\pi}{3 M_{\rm Pl}} \frac{m^2}{x^2} ~,
\end{eqnarray}
where $M_{\rm Pl} = 2.4 \times 10^{18}$ GeV is the reduced Planck mass.
The relativistic degrees of freedom of the thermal bath, $g_*$ and $g_{*s}$,
should be treated as a function of the temperature
for deriving the correct dark matter abundance in our calculation.

\begin{figure}[t]
\begin{center}
\scalebox{0.55}{\includegraphics*{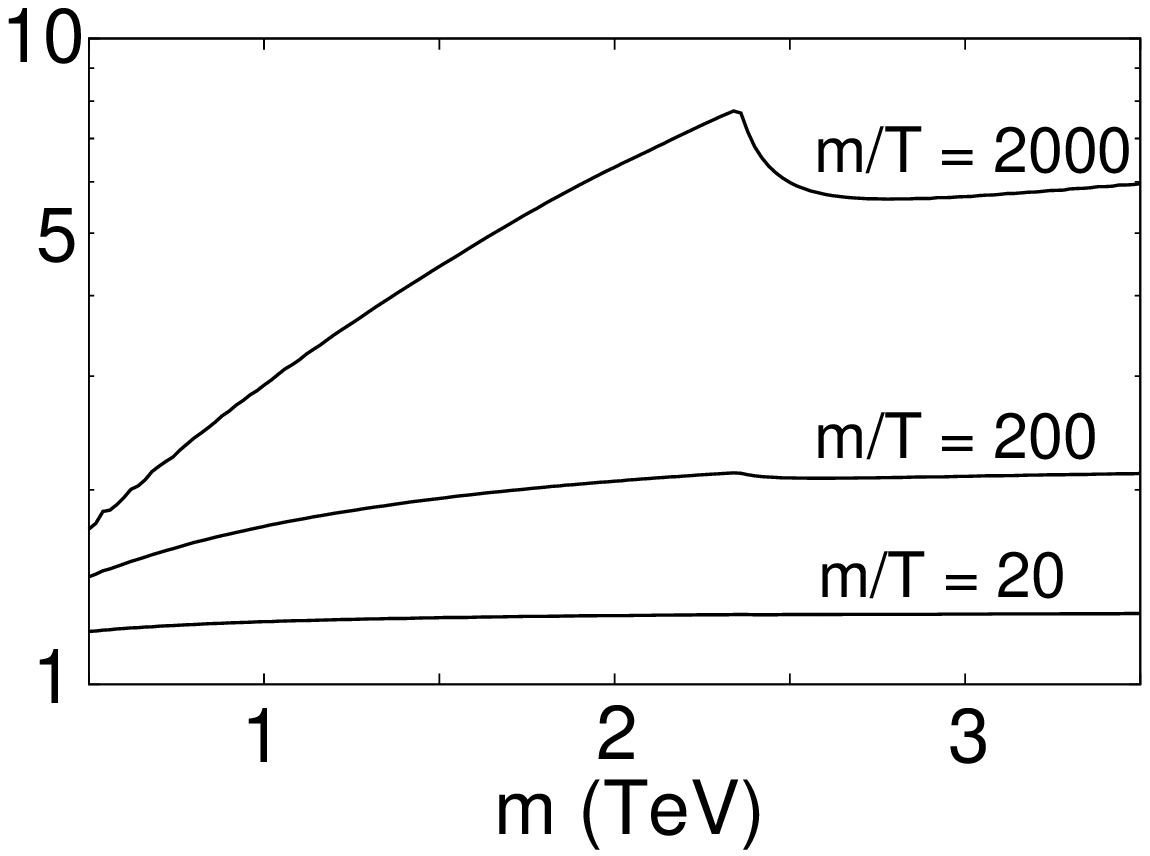}}
\put(-120,138){$\langle \sigma_{\rm eff} \rangle /
\langle \sigma_{\rm eff} \rangle_{\rm Tree}$}
\qquad
\scalebox{0.55}{\includegraphics*{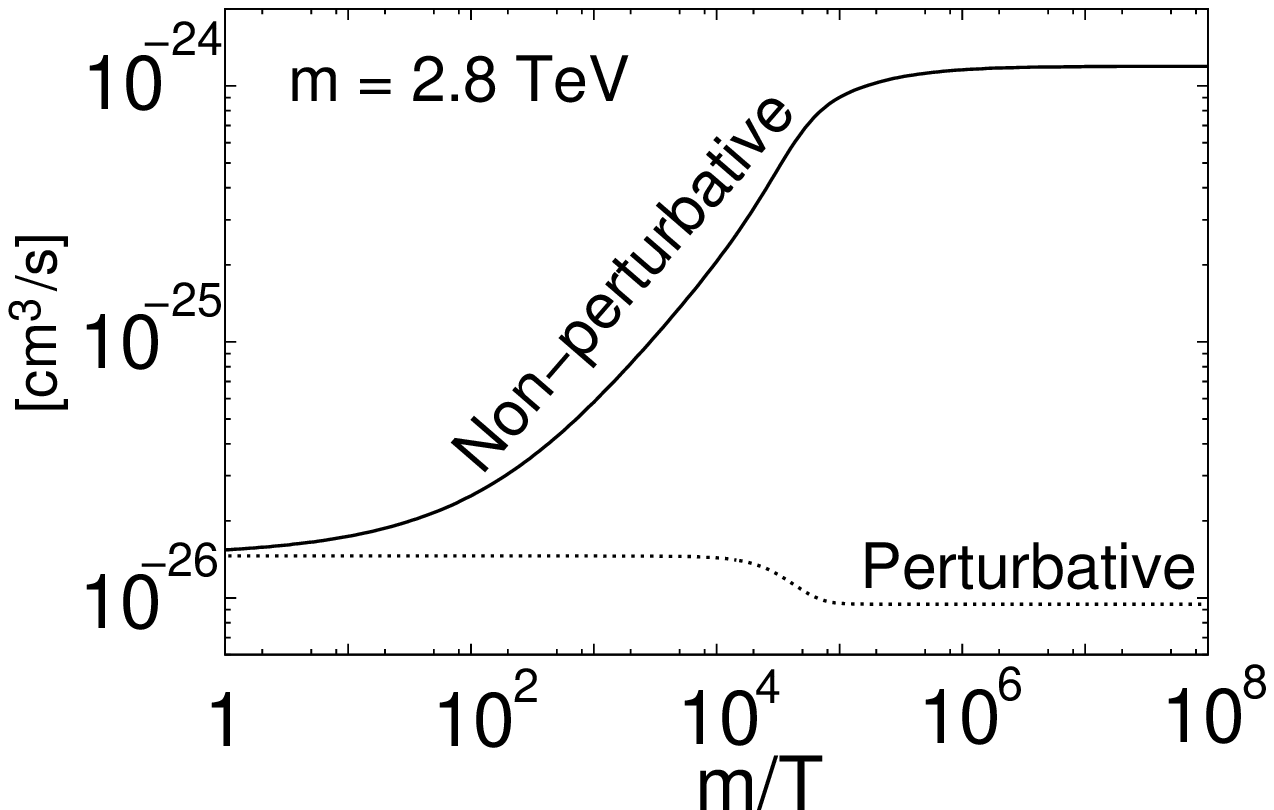}}
\put(-177,134){\small Thermal averaged cross section, $\langle \sigma_{\rm eff} \rangle$}
\caption{\small Mass dependence of  averaged cross section,
  $\langle \sigma_{\rm eff} \rangle$, normalized by the perturbative
  one, $\langle \sigma_{\rm eff} \rangle_{\rm Tree}$, for $m/T =
  20,~200,~2000$ (left figure), and temperature dependence with $m
  = 2.8$~TeV (right figure).  The perturbative result is also shown as
  a dotted line for comparison. Here, mass difference between
$\DM$ and $\tilde \chi^\pm$ is  $\delta m \simeq$ 0.17 GeV.}
\label{CS}
\end{center}
\end{figure}

The most important quantity to determine the abundance is the thermally
averaged effective annihilation cross section $\langle\sigma_{\rm eff}v\rangle$
in Eq.~(\ref{eq:Boltzmann}), defined as
\begin{eqnarray}
 \langle\sigma_{\rm eff}v\rangle
 &=&
 \sum_{i,j}\langle\sigma_{ij}v\rangle\frac{4}{g_{\rm eff}^2(x)} 
 (1 + \Delta_i)^{3/2}(1 + \Delta_j)^{3/2}
 \exp[-x(\Delta_i + \Delta_j)]~,
 \nonumber \\
 \langle\sigma_{ij}v\rangle
 &=&
 \left(\frac{m}{4\pi T}\right)^{3/2}
 \int 4\pi v^2dv~\left(\sigma_{ij}v\right)
 \exp\left(-\frac{mv^2}{4T}\right)~,
 \label{effective CS}
\end{eqnarray}
where $i,j = \DM$, $\CP$ and $\CPC$, $\Delta_{\CP} = \Delta_{\CPC} =
\delta m / m$, $\Delta_{\DM} = 0$, and $\sigma_{ij}$ is the
annihilation cross section between $i$ and $j$.  In Fig.~\ref{CS}, the
enhancement ratio of the averaged cross section, $\langle\sigma_{\rm
  eff}v\rangle$, to that in the perturbative calculation is shown as a
function of $m$ with fixed $m/T = 20, ~200, ~2000$ (left figure) and
as a function of $T$ with fixed $m = 2.8$ TeV (right figure). For
comparison, the cross section in a perturbative calculation is also
shown as a dotted line in the right figure.  Note that the
perturbative cross section is constant in time.  The little drop at $x
\sim 10^5$ is due to the decoupling of $\tilde \chi ^\pm$.  In the
calculation of the cross section, we used the running gauge coupling
constant at the $m$ and $m_Z$ in the absorptive terms and the
potentials, respectively.

In these figures, large enhancement of the cross section is found due
to the non-perturbative effect when $m$ is larger than $\sim 1$~TeV.
A significant enhancement is shown at $m\sim$ 2.4 GeV. This originates
in the bound state composed of $\DM\DM$ and $\CP\CPC$ pairs
\cite{Hisano:2004ds,Hisano:2003ec}.
The enhancement by the non-perturbative effect
is more significant for lower temperature.
Since $\DM$ and $\CP$ are more non-relativistic for lower temperature,
the long-range force acting between these particles strongly
modifies their wave functions and alters the cross section significantly.

\begin{figure}[t]
\begin{center}
\scalebox{0.55}{\includegraphics*{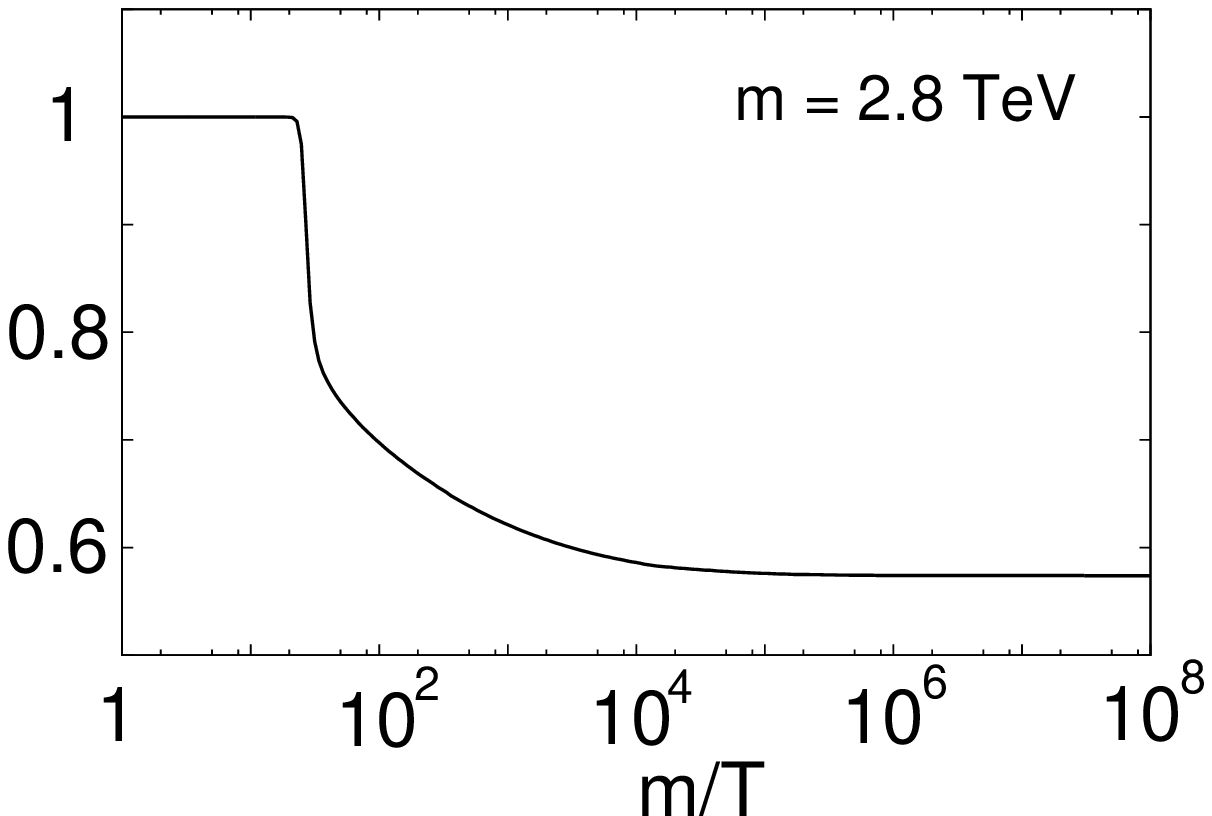}}
\put(-115,136){$Y / Y_{\rm Tree}$}
\qquad
\scalebox{0.55}{\includegraphics*{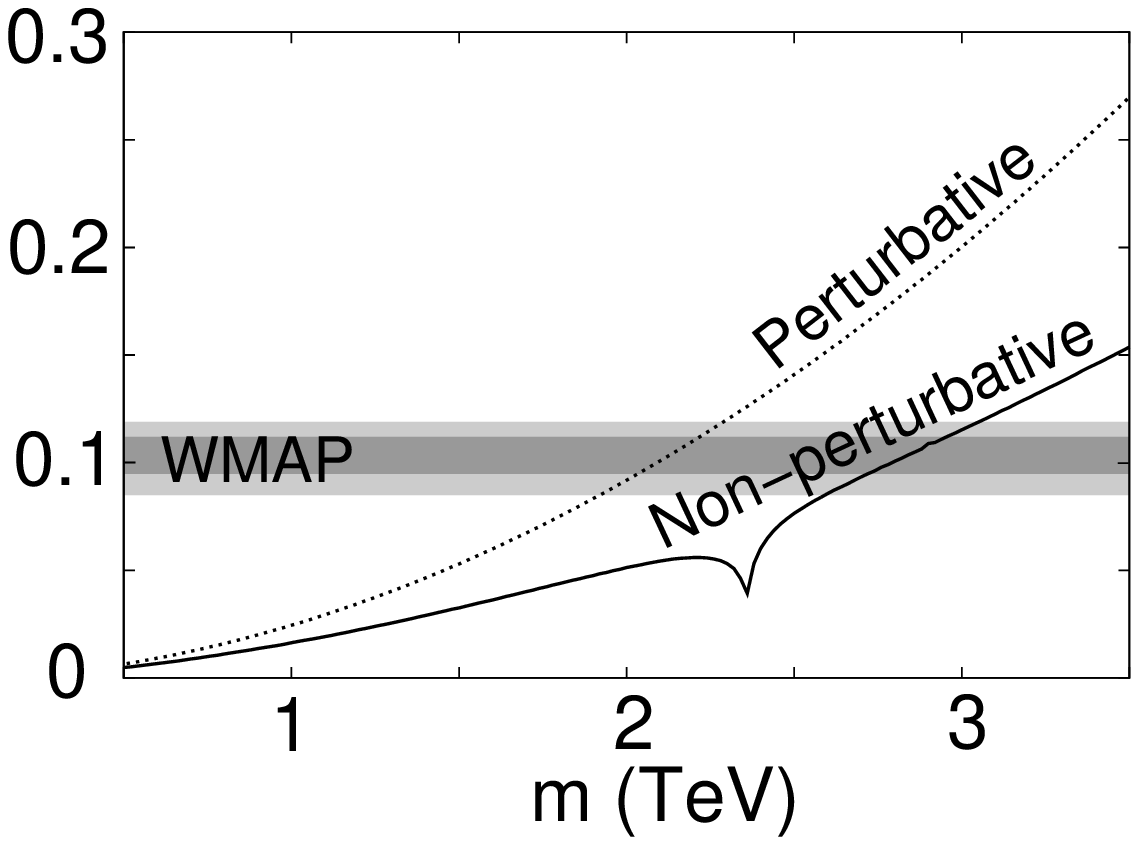}}
\put(-165,136){Thermal relic abundance, $\Omega_{\rm DM} h^2$}
\caption{\small Ratio of yield with the non-perturbative effect to
  that in the perturbative calculation (left figure). Wino-like
  neutralino mass is fixed 2.8 TeV. Thermal relic abundance of
  the dark matter in the current universe as a function of 
  wino-like neutralino mass (right figure).  Allowed regions by
  the WMAP at 1(2)~$\sigma$ levels are also shown as the dark (light)
  shaded area.}
\label{Omega}
\end{center}
\end{figure}

Since the averaged cross section depends on temperature in a
non-trivial way as shown in Fig.~\ref{CS}, we should integrate the
Boltzmann equation numerically.  After calculating the present value
of the yield, $Y_0$, by the integration, we obtain the dark matter
mass fraction in the current universe through the relation
$\Omega_{DM} h^2 = m s_0 Y_0 h^2/\rho_c$, where $\rho_c$ is the
critical density, $\rho_c = 1.05 \times 10^{-5} h^2 ~{\rm GeV
  cm}^{-3}$ ($h = 0.73^{+0.04}_{-0.03}$) \cite{PDG}, and $s_0$ is the
entropy density of the present universe.  

The result is shown in Fig.~\ref{Omega}.  In the left figure, the
ratio of the yield with the enhancement to one without the enhancement
(perturbative result) is shown as a function of temperature.  The mass
of the wino-like neutralino is assumed $ m = 2.8$~TeV.  The
enhancement of the cross section at the departure from the equilibrium
decreases the abundance by $20-30\%$, and it leads to quick deviation
of the yield from the perturbative one after decoupling.  Since the
annihilation cross section is increased for lower temperature by the
Sommerfeld enhancement, the sudden freeze-out phenomenon on the yield
does not occur and the yield continues to be reduced by the
non-perturbative effect even for $x > 100$ compared to the
perturbative one.  The resultant dark matter abundance is reduced by
50\% compared to the perturbative result.

In the right figure, the relic abundance of the dark matter in the
present universe is shown as a function of $m$ in terms of $\Omega
h^2$.  The allowed regions by the WMAP at 1 and 2 $\sigma$ are also shown
as shaded areas in this figure.  We found that the mass in the
wino-like neutralino dark matter consistent with the observation is
shifted by 600 GeV due to the non-perturbative effect and the
wino-like neutralino mass consistent with WMAP results turns out to be
2.7~TeV $\lesssim m \lesssim$ 3.0~TeV.

\section{Summary and discussion}

In this letter, we have pointed out the thermal relic abundance of
dark matter, which is SU(2)$_L$ non-singlet and has a much larger mass
than that of the weak gauge bosons, can be strongly reduced by the
non-perturbative effect.  We have investigated the non-perturbative
effect on the relic abundance of wino-like neutralino as an example.
Compared with the perturbative result, this effect reduces the
abundance by about 50\% and increases the mass of the wino-like
neutralino dark matter consistent with the observation by about 600
GeV.  As a result, the thermal relic abundance of the wino-like
neutralino dark matter is consistent with observed abundances when
2.7~TeV $\lesssim m \lesssim$ 3.0~TeV.

The non-perturbative effect can change relic abundances of other dark
matter candidates with SU(2)$_L$ charge and heavy mass, such as
higgsino-like neutralino.  The non-perturbative effect on the thermal
relic abundance of higgsino-like neutralino is expected to be
roughly 10\%, since winos are triplet under the SU(2)$_L$
gauge group, while higgsinos are SU(2)$_L$ doublet.  Therefore,
non-perturbative effect may change the abundance by O(10\%) for other
SU(2)$_L$ doublet candidates for the dark matter.  A detailed analysis
of this subject is studied elsewhere.

The Sommerfeld enhancement occurs reasonably for particles with
electric charge.  In fact the non-perturbative effect through photon
exchanges for charged particle annihilation can change the abundance
by about 10\%.  Therefore, one may think that the relic abundance of
dark matter will be changed in the stau coannihilation region or in
the case that gravitino is the lightest supersymmetric particle and it
is produced through decay of stau.  However, the non-perturbative
effect for stau does not change the dark matter abundance.  In stau
annihilation, $\sigma (\tilde \tau ^+ \tilde \tau ^-) \simeq \sigma
(\tilde \tau ^+ \tilde \tau ^+)$ and the Sommerfeld enhancement is
positive (negative) for the former (latter) process.  Hence, the
enhancement is almost canceled and the change of the abundance is 1\%
at most.

Kaluza-Klein (KK) right-handed leptons ($E^{(1)}$) in UED models are
highly degenerate with the lightest KK particle in mass and
$\sigma(E^{(1)} \bar E ^{(1)}) > \sigma(E^{(1)} E ^{(1)})$
\cite{UED2,Kakizaki:2005uy}.
Hence, the enhancement could be expected to change the
abundance of the KK dark matter.  The change of the abundance is within 4\%
since $\sigma(E^{(1)} \bar E ^{(1)})$ contributes to the effective
annihilation cross section by 40\% at most.

Finally, we comment on the non-perturbative effect of colored particles,
such as gluino.  The enhancement for colored particles are very
effective \cite{Baer:1998pg}.  However, it may be very complicated due
to the existence of the QCD phase transition, which is discussed in
Ref.~\cite{Arvanitaki:2005fa}.  Furthermore, colored particles can not
be candidates for the dark matter and are not expected to be degenerate
with a dark matter particle in mass due to large radiative corrections
by the strong interaction.  Hence, this subject is beyond the scope of
this letter.

\section*{Acknowledgments}

This work is supported in part by the Grant-in-Aid for Science
Research, Ministry of Education, Science and Culture, Japan
(No.1803422 and 15540255 for JH, 16081211 for SM and 18840011 for MS).
Also, the work of MN is supported in part by JSPS.

\end{document}